\documentclass[aps,prl,showpacs,twocolumn,groupedaddress]{revtex4}
\usepackage{graphicx}
\begin{document}

\title{Liquid heat capacity in the approach from the solid state: anharmonic theory}
\author{Dima Bolmatov$^{1}$\footnote{e-mail: d.bolmatov@qmul.ac.uk}}
\author{Kostya Trachenko$^{1,2}$}
\address{$^1$ School of Physics, Queen Mary University of London, Mile End Road, London, E1 4NS, UK}
\address{$^2$ South East Physics Network}
\begin{abstract}
Calculating liquid energy and heat capacity in general form is an open problem in condensed matter physics. We develop a recent approach to liquids from the solid state by accounting for the contribution of anharmonicity and thermal expansion to liquid energy and heat capacity. We subsequently compare theoretical predictions to the experiments results of 5 commonly discussed liquids, and find a good agreement with no free fitting parameters. We discuss and compare the proposed theory to previous approaches.
\end{abstract}

\maketitle

\section{Introduction}
Among three basic states of matter (solid, liquid, gas), liquids are least understood from the theoretical point of view. The inter-atomic, or inter-molecular, interactions in a liquid are strong, and therefore strongly affect the liquid energy. At the same time, the interactions are system-specific, hence the calculation of liquid energy requires the explicit knowledge of the interactions. For this reason, it is argued \cite{landau} that no general expressions for liquid energy can be obtained, in contrast to gases and solids.

According to current theoretical understanding, liquids are strikingly different from both gases and solids. Indeed, small atomic displacements in a solid make it possible to expand the energy in terms of phonons and obtain general expressions for solid energy. On the other hand, this has largely considered to be impossible to do in liquids where atomic displacements are large. Similarly, small interactions in gases make it possible to treat interactions as a perturbation and obtain general expressions for the corrections to the system energy. On the other hand, this has not been useful for liquids with strong interactions and solid-like densities. An apt summary of this state of affairs, attributed to Landau, is that liquids ``have no small parameter''. Perhaps for this reason, liquid heat capacity is not, or is barely, mentioned in statistical physics textbooks as well as books dedicated to liquids \cite{landau,frenkel,ziman,yip,march,hansen}. This observation is shared by Granato \cite{granato}, who further comments on the challenge faced by teachers to discuss liquid heat capacity in class.

The experimental behavior of liquid heat capacity is interesting. The heat capacity per atom {\it decreases} from approximately $3k_{\rm B}$ per atom at the melting point to about $2k_{\rm B}$ at high temperature, as witnessed by the data of many simple liquids \cite{grimvall,wallace}. Similar behavior is also seen in complex liquids \cite{dexter}. The decrease of heat capacity was observed in molecular dynamics simulations \cite{md,md1}. Liquid heat capacity has also been discussed by considering liquid potential energy landscape with inter-valley motions \cite{wallace}.

Theoretically, liquids have been viewed to occupy an intermediate state between gases and solids. Liquids are fluid, and therefore might intuitively appear closer to gases in terms of their properties. On the other hand, unless they are close to critical point, liquids have solid-like densities and can support shear waves at high frequencies. The question of how best to treat real dense liquids, i.e. approach them from the gas or solid state, has a long history \cite{frenkel}.

Historically, liquids have predominantly been approached from the gas state, by developing schemes to calculate the interaction energy in addition to gas kinetic energy \cite{landau,frenkel,ziman,yip,march,hansen}. This approach relies on the knowledge of interatomic interactions as well as correlation functions \cite{born,henders}. These are not generally available, apart from simple model systems, as discussed below in more detail. When they are available, it is not apparent how this approach explains the experimental decrease of heat capacity from $3k_{\rm B}$ to $2k_{\rm B}$ at high temperature.

An alternative approach to liquids was pioneered by Frenkel \cite{frenkel}, and is based on liquid relaxation time $\tau$. $\tau$ is the average time between two consecutive local structural rearrangements in a liquid at one point in space. In this picture, a liquid is approached from the solid state because locally, liquid structure remains unchanged, i.e. the same as in a solid, during time shorter than $\tau$. An important advantage of this approach is that strong interactions are included in the consideration from the outset. This is in contrast to the approach from the gas state that attempts to calculate the interaction energy from correlation functions and interatomic interactions.

Frenkel noted that as $\tau$ continuously increases (if crystallization is avoided) beyond the experimental time scale, a liquid becomes a solid for practical purposes. Hence, a solid is different from a liquid only quantitatively but not qualitatively. Frenkel subsequently stated: ``...the classification of condensed bodies into solids and liquids... has a relative meaning convenient for practical purposes but devoid of scientific value'' \cite{frenkel}. Novel for that time, this view might perhaps come across as somewhat unusual even at present. We suggest that a possible reason for this is that solid-like properties of liquids have not become apparent in the traditional approach to liquids from the gas state. On the other hand, we propose here that as far as their energy and heat capacity are concerned, liquids can be understood on the basis of their solid-like properties, consistent with Frenkel's general idea.

We have recently proposed how liquid energy can be calculated in Frenkel's approach, by relating liquid energy to $\tau$ \cite{prb}. We have found good agreement of liquid heat capacity between theoretical and experimental data for mercury. Two important questions subsequently arise. First, how well does the theory work for a larger number of liquids? Second, the proposed theory was harmonic. On other hand, anharmonic effects are known to be particularly large in liquids, where the coefficients of thermal expansion considerably exceed those in solids \cite{prb}. Hence, it is important to extend the theory to the anharmonic case and compare it with experiments.

In this paper, we propose how to include the anharmonic effects in the approach to liquids from the solid state. We subsequently calculate heat capacity from both harmonic and anharmonic theory, and compare the results with experimental data for 5 commonly discussed liquids. We find good agreement between theoretical predictions and experimental data with no free fitting parameters. Finally, we discuss and compare the proposed theory to the previous approaches to liquids.

\section{Harmonic theory}

We begin our discussion with the work of Frenkel \cite{frenkel}, who provided a microscopic description of Maxwell phenomenological viscoelastic theory of liquid flow \cite{max}, by introducing liquid relaxation time $\tau$. As mentioned earlier, $\tau$ is the average time between two consecutive atomic jumps in a liquid at one point in space. Each jump can approximately be viewed as a jump of an atom from its neighboring cage into a new equilibrium position, with subsequent cage relaxation. These atomic jumps, or local relaxation events (LREs), give a liquid its ability to flow. $\tau$ is a fundamental flow property of a liquid, and is directly related to liquid viscosity $\eta$ as $\eta=G_{\infty}\tau$ \cite{max,frenkel}, where $G_{\infty}$ is the instantaneous shear modulus. On temperature decrease, $\tau$ increases by many orders of magnitude, reaching $10^2-10^3$ s at which point, by convention, a liquid forms a solid glass because LREs stop operating on a typical experimental time scale.

In Frenkel's theory, motion of an atom in a liquid consists of two types: vibrational motion around an equilibrium position as in a solid, with Debye vibration period of about $\tau_{\rm D}=0.1$ ps and diffusional motion between two neighboring positions during time $\tau$. Therefore, if the observation time is smaller than $\tau$, the local structure of a liquid does not change, and is the same as that of a solid glass. In this picture, Frenkel realized that a liquid should maintain solid-like shear waves, similarly to those existing in a solid, at all frequencies $\omega$ larger than $\frac{1}{\tau}$ \cite{frenkel}. This prediction was later confirmed experimentally \cite{copley,grim,pilgrim,burkel,rec-review}. Longitudinal waves, associated with density fluctuations, are considered to be unaffected in this picture, apart from different dissipation laws for $\omega<\frac{1}{\tau}$ and $\omega>\frac{1}{\tau}$ \cite{frenkel}.

Basing on the ability to support high-frequency shear waves, we have calculated liquid energy as follows \cite{prb}. In Frenkel's picture, liquid energy is

\begin{equation}
E=E_v+E_d
\end{equation}

\noindent where $E_v$ and $E_d$ are the energies of vibration and diffusion, respectively.

The vibration energy consists of the energy of one longitudinal mode and two shear modes with frequency $\omega>\frac{1}{\tau}$. Then, $E_v$ can be written as $E_v=K_l+P_l+K_s(\omega>\frac{1}{\tau})+P_s(\omega>\frac{1}{\tau})$ where $K$ and $P$ correspond to kinetic and potential terms, respectively. Here, the absence of shear modes with frequency $\omega<\frac{1}{\tau}$ is expressed as $K_s(\omega<\frac{1}{\tau})=P_s(\omega<\frac{1}{\tau})=0$, implying the absence of restoring forces for low-frequency vibrations. Similarly, $E_d$ can be written as $E_d=K_d+P_d$, where $K_d$ and $P_d$ are respective kinetic and potential terms, giving

\begin{equation}
E=K_l+P_l+K_s(\omega>\frac{1}{\tau})+P_s(\omega>\frac{1}{\tau})+K_d+P_d
\label{en}
\end{equation}

We now approach a liquid from the solid state (glass) where diffusion is absent so that $E_d=0$ and $E=E_v$. At certain temperature, shear waves with frequency $\omega<\frac{1}{\tau}$ disappear. The oscillatory motion at frequency smaller than $\omega<\frac{1}{\tau}$ is substituted by diffusional motion taking place during time $\tau$. The latter motion can be thought of as a ``slipping'' motion at low frequency because the restoring force for low-frequency vibrations becomes small (see Figure 1). We note that forces experienced by diffusing atoms are of the same nature as those experienced by vibrating atoms, and are defined by interatomic interactions that give rise to both $P_d$ and $P_s$. Therefore, the relative smallness of $P_s$ at small frequency implies the smallness of $P_d$ as compared to $P_s(\omega>\frac{1}{\tau})$ (as well as to $P_l$). In other words, if the interaction energy of diffusing atoms, $P_d$, were large and comparable to $P_s(\omega>\frac{1}{\tau})$, this would imply strong restoring forces and consequently the existence of low-frequency vibrations. Therefore, $P_d$ can be neglected in Eq. (\ref{en}). This is the only approximation in the theory \cite{prb}. We further note that the energy of diffusion becomes unimportant at low temperature \cite{gla-tr}. Indeed, because the time that an atom spends in the transitory diffusing state is approximately $\tau_{\rm D}$, the relative number of diffusing atoms, $\frac{N_d}{N}$, is equal to the jump probability, $\frac{\tau_{\rm D}}{\tau}$, where $N_d$ and $N$ are the number of diffusing atoms and the total number of atoms, respectively. When $\tau$ significantly exceeds $\tau_{\rm D}$, the number of diffusing atoms and, therefore, diffusing energy, become small and can be ignored.

\begin{figure}
\begin{center}
{\scalebox{0.85}{\includegraphics{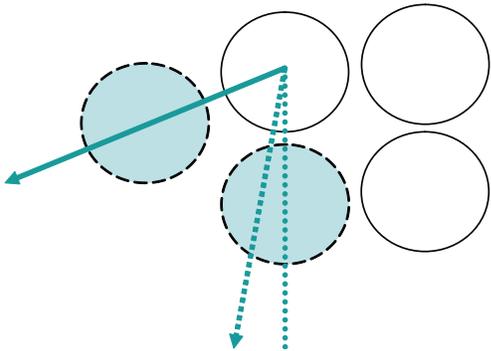}}}
\end{center}
\caption{As temperature increases, the low-frequency oscillatory shear motion of the highlighted atom between the two dashed lines is lost, as the restoring force for low-frequency shear vibrations becomes weak. Instead, the atom ``slips'' and diffuses to another position defined by the solid arrow. Schematic illustration.}
\end{figure}

We now note that the sum of all kinetic terms in Eq. (\ref{en}) gives the total kinetic energy of the liquid, $K$. Indeed, $K=\frac{3NT}{2}$ regardless of how the motion partitions into vibrational and diffusional motion. Then, Eq. (\ref{en}) becomes

\begin{equation}
E=K+P_l+P_s(\omega>\frac{1}{\tau})
\label{en1}
\end{equation}

It is convenient to re-write Eq. (\ref{en1}) using the equipartition theorem that implies $P_l=\frac{E_l}{2}$, $P_s(\omega>\frac{1}{\tau})=\frac{E_l(\omega>\frac{1}{\tau})}{2}$ and $K=K_l+K_s=\frac{E_l}{2}+\frac{E_s}{2}$, where we noted that liquid kinetic energy is the same as in the solid ($K=\frac{3NT}{2}$), and can therefore be written as a sum of kinetic terms related to longitudinal and shear waves. Then, $E$ in Eq. (\ref{en1}) becomes $E=E_l+\frac{E_s}{2}+\frac{E_s(\omega>\frac{1}{\tau})}{2}$. For subsequent calculations, it is convenient to further write $E_s=E_s(\omega<\frac{1}{\tau})+E_s(\omega>\frac{1}{\tau})$, where the two terms refer to their solid-state values. Then, $E$ becomes finally

\begin{equation}
E=E_l+E_s(\omega>\frac{1}{\tau})+\frac{E_s(\omega<1/\tau)}{2}
\label{en2}
\end{equation}

The first two terms can be calculated in the same way as is done in the harmonic theory of solids \cite{landau}, except we separate longitudinal and shear waves in the partition function and account for shear waves with frequency $\omega>\frac{1}{\tau}$ only. Let $Z_2$ be associated with the first two terms in Eq. (\ref{en2}). Then, $Z_2$ is:

\begin{eqnarray}
Z_2=(2\pi\hbar)^{-N^\prime}\int\exp\left(-\frac{1}{2T}\sum\limits_{i=1}^N(p_i^2+\omega_{li}^2q_i^2)\right) dpdq\\\times
\int\exp\left(-\frac{1}{2T}\sum\limits_{\omega_{si}>\omega_0}^{2N}(p_i^2+\omega_{si}^2q_i^2)\right)dpdq\nonumber
\label{part}
\end{eqnarray}

\noindent where $\omega_0=\frac{1}{\tau}$, $\omega_{li}$ and $\omega_{si}$ are frequencies of longitudinal and shear waves, $N$ is the number of atoms and $N^\prime$ is the number of phonon states that include longitudinal waves and transverse waves with frequency $\omega>\frac{1}{\tau}$. Here and below, $k_{\rm B}=1$.

Integrating, we find

\begin{equation}
Z_2=T^N\left(\prod\limits_{i=1}^N\hbar\omega_{li}\right)^{-1}T^{N_1}\left(\prod\limits_{\omega_{si}>\omega_0}^{2N}\hbar\omega_{si}\right)^{-1}
\label{part}
\end{equation}

\noindent where $N_1$ is the number of transverse modes with $\omega>\omega_0$.

In the harmonic approximation, frequencies $\omega_{li}$ and $\omega_{ti}$ are considered to be temperature-independent, in contrast to anharmonic case discussed in the next section. Then, Eq. (\ref{part}) gives the liquid energy in harmonic approximation $E=T^2\frac{d}{dT}\ln Z=NT+N_1T$. $N_1$ can be calculated using the quadratic density of states in the Debye model, as is done in solids \cite{landau}. The density of states of shear modes is $g_t(\omega)=\frac{6N}{\omega_{mt}^3}\omega^2$, where $\omega_{mt}$ is Debye frequency of shear modes ($\omega_{mt}\approx\omega_{\rm D}$) and we have taken into account that the number of shear waves is $2N$. Then, $N_1=\int\limits_{\omega_0}^{\omega_{mt}}g_t(\omega)d\omega=2N\left(1-\left(\frac{\omega_0}{\omega_{mt}}\right)^3\right)$.

To calculate the last term in Eq. (\ref{en2}), we note that similarly to $E_t(\omega>1/\tau)=N_1 T$, $E_t(\omega<1/\tau)=N_2 T$, where $N_2$ is the number of shear modes with $\omega<\omega_0$. Because $N_2=2N-N_1$, $N_2=2N\left(\frac{\omega_0}{\omega_{mt}}\right)^3$. Then, $E=(N+N_1+\frac{N_2}{2})T$, giving finally

\begin{equation}
E=NT\left(3-\left(\frac{\omega_0}{\omega_{\rm D}}\right)^3\right)
\label{harmo}
\end{equation}

According to Eq. (\ref{harmo}), the liquid energy is $3NT$ as in a solid when $\tau\gg\tau_{\rm D}$. This gives the heat capacity per atom, $c_v=\frac{1}{N}\frac{{\rm d}E}{{\rm d} T}$, $c_v=3$, consistent with experimental results \cite{grimvall,wallace}. When $\tau\rightarrow\tau_{\rm D}$ at high temperature, Eq. (\ref{harmo}) predicts $c_v=2$, as in the experimental data \cite{grimvall,wallace}.

\section{Anharmonic theory}

In the harmonic approximation, temperature dependence of frequencies in Eq. (\ref{part}) is ignored. This may be a good approximation for phonons in solids at either low temperature or in systems with small thermal expansion coefficient $\alpha$. In liquids, on the other hand, anharmonicity and associated thermal expansion are large \cite{anderson}, and need to be taken into account. In this case, applying $E=T^2\frac{d}{dT}\ln Z$ to Eq. (\ref{part}) gives

\begin{equation}
E_2=NT-T^2\sum\limits_{i=1}^N\frac{1}{\omega_{li}}\frac{{\rm d}\omega_{li}}{{\rm d}T}+N_1T-T^2\sum\limits_{i=1}^{N_1}\frac{1}{\omega_{si}}\frac{{\rm d}\omega_{si}}{{\rm d}T}
\label{anharm}
\end{equation}

\noindent which corresponds to the first two terms in Eq. (\ref{en2}).

The anharmonic effects lead to the decrease of frequencies with temperature, resulting in system softening \cite{anderson}. We therefore need to calculate $\frac{{\rm d}\omega_{li}}{{\rm d}T}$ and $\frac{{\rm d}\omega_{si}}{{\rm d}T}$ in Eq. (\ref{anharm}). These can be calculated in the Gr\"{u}neisen approximation, by introducing the Gr\"{u}neisen parameter $\gamma=-\frac{V}{\omega}\left(\frac{\partial\omega_i}{\partial V}\right)_T$, where $\omega_i$ are frequencies in harmonic approximation \cite{gla-tr}. $\gamma$ features in the phonon pressure $P_{ph}=-\left(\frac{\partial F}{\partial V}\right)_T$, where $F$ is free energy in the harmonic approximation. $F=-T\ln Z$ can be calculated from Eq. (\ref{part}), giving

\begin{equation}
F_2=T\sum\limits_{i=1}^N\ln\frac{\hbar\omega_{li}}{T}+T\sum\limits_{i=1}^{N_1}\ln\frac{\hbar\omega_{si}}{T}
\end{equation}

Calculating $P_{ph}=-\left(\frac{\partial F}{\partial V}\right)_T$ and introducing the above Gr\"{u}neisen parameter gives $P_{ph}=\frac{\gamma T}{V}(N+N_1)$. Then, the bulk modulus due to the (negative) phonon pressure is $B_{ph}=V\frac{\partial P}{\partial V}=-\frac{\gamma T}{V}(N+N_1)$, giving $\left(\frac{\partial B_{ph}}{\partial T}\right)_v=-\frac{\gamma}{V}(N+N_1)$. We now use the macroscopic definition of $\gamma$, $\gamma=\frac{V\alpha B}{C_v}$ \cite{anderson}. Here, $B=B_0+B_{ph}$ is the total bulk modulus, $B_0$ is zero-temperature bulk modulus, $\alpha$ is the coefficient of thermal expansion and $C_v$ is constant-volume heat capacity. For $C_v$, we use its harmonic value, $C_v=N+N_1$ from Eq. (\ref{anharm}), because  $\frac{{\rm d}\omega_{li}}{{\rm d}T}$ and $\frac{{\rm d}\omega_{si}}{{\rm d}T}$ in Eq. (\ref{anharm}) already enter as quadratic anharmonic corrections. Then, $\left(\frac{\partial B_{ph}}{\partial T}\right)_v=-\alpha(B_0+B_{ph})$. For small $\alpha T$, as is often the case in experiments, this implies $B\propto -T$, consistent with the experimental data \cite{anderson}.

We note that experimentally, $B$ linearly decreases with $T$ at both constant volume and constant pressure \cite{anderson}. The decrease of $B$ with $T$ at constant volume is due to the intrinsic anharmonicity related to the softening of interatomic potential at large vibrational amplitudes; the decrease of $B$ at constant pressure has an additional contribution from thermal expansion.

Assuming $\omega^2\propto B_0+B_{ph}$ and combining it with $\left(\frac{\partial B_{ph}}{\partial T}\right)_v=-\alpha(B_0+B_{ph})$ from above gives $\frac{1}{\omega}\frac{{\rm d}\omega}{{\rm d}T}=-\frac{\alpha}{2}$ \cite{gla-tr}. Putting it in Eq. (\ref{anharm}) gives:

\begin{equation}
E_2=(N+N_1)\left(T+\frac{\alpha T^2}{2}\right)
\label{anharm1}
\end{equation}

The last term in Eq. (\ref{en2}), $\frac{E_s(\omega<1/\tau)}{2}$, can be calculated in the same way, giving $\frac{1}{2}N_2\left(T+\frac{\alpha}{2}T^2\right)$, where $N_2$ is the number of shear modes with $\omega<\omega_0$ defined in the previous section. Adding this term to Eq. (\ref{anharm1}) and using $N_1$ and $N_2$ calculated in the previous section gives the anharmonic liquid energy:

\begin{equation}
E=N\left(T+\frac{\alpha T^2}{2}\right)\left(3-\left(\frac{\omega_0}{\omega_{\rm D}}\right)^3\right)
\label{anharm2}
\end{equation}

At low temperature when $\tau$ exceeds $\tau_{\rm D}$, Eq. (\ref{anharm2}) gives $E=3N(T+\frac{\alpha T^2}{2})$, and $c_v$ is

\begin{equation}
c_v=3(1+\alpha T)
\label{sol}
\end{equation}

Eq. (\ref{sol}) explains why experimental $c_v$ of liquids exceed the Dulong-Petit value just above the melting point \cite{grimvall,wallace}. In the next section, we compare the calculated and experimental $c_v$ in the entire temperature range.

\section{Comparison with experimental data}

We now compare the predictions of Eqs. (\ref{harmo}) and (\ref{anharm2}) with the experimental $c_v$. Simple liquids \cite{grimvall,wallace} offer a good test case for a number of reasons. First, they are frequently discussed in the literature. Consequently, the data of both $c_v$ and viscosity are available for simple liquids, partially because their melting temperatures are fairly low, simplifying the measurements. Second, these liquids become low-viscous close to the melting point, in contrast to, for example, liquid B$_2$O$_3$, SiO$_2$ and so on. Consequently, the decrease of $c_v$, predicted by Eqs. (\ref{harmo}) and (\ref{anharm2}) to operate around $\tau\approx\tau_{\rm D}$, can be observed in a convenient and accessible temperature range. Third, it is important to see that our theory works in a wide temperature range. We have taken experimental $c_v$ for liquid Hg, In, Rb, Cs and Sn where $c_v$ has been measured in a wide range of temperatures, from 200 to 1200 K.

To the best of our knowledge, apart from liquid Hg, this is the first attempt to analytically calculate $c_v$ of the above liquids and compare them with experiments.

To calculate $c_v$ from the theory, we write $\tau=\frac{1}{\omega_0}$ in terms of liquid viscosity using the Maxwell relationship $\tau=\frac{\eta}{G_{\infty}}$. Then, harmonic and anharmonic $c_v$ in Eq. (\ref{harmo}) and Eq. (\ref{anharm2}) are given by the two equations below:

\begin{equation}
c_v=\frac{\rm d}{{\rm d} T}\left(T\left(3-\left(\frac{\tau_{\rm D}G_{\infty}}{\eta}\right)^3\right)\right)
\label{10}
\end{equation}

\begin{equation}
c_v=\frac{\rm d}{{\rm d} T}\left(\left(T+\frac{\alpha T^2}{2}\right)\left(3-\left(\frac{\tau_{\rm D}G_{\infty}}{\eta}\right)^3\right)\right)
\label{11}
\end{equation}

We have taken viscosity data from Refs \cite{eta1,eta2,eta3} and fitted them to the form of the Vogel-Fulcher-Tammann law, $\eta=\eta_0\exp\left(\frac{A}{T-T_0}\right)$. The data were subsequently extrapolated on the temperature range of experimental $c_v$ because the experimental $c_v$ was measured in a wider temperature range than $\eta$, and used $\eta$ in Eqs. (\ref{10},\ref{11}) to calculate $c_v$. We note that $c_v$ in Eqs. (\ref{10},\ref{11}) depends on both $\eta$ and $\frac{{\rm d}\eta}{{\rm d} T}$, and is therefore sensitive to the viscosity fit. We find that as long as the VFT fit has physical values of parameters (e.g. positive $T_0$), a reasonable agreement between the calculated and experimental $c_v$ is found, as discussed below in more detail.

Notably, Eqs. (\ref{10}) and (\ref{11}) have no free fitting parameters. Indeed, a single parameter in Eq. (\ref{10}), $\tau_{\rm D}G_{\infty}$, is defined by the liquid properties. Eq. (\ref{11}) has an extra parameter $\alpha$, which is equally governed by liquid properties. In practice, however, the values of $\tau_{\rm D}$ and $G_{\infty}$ are not known precisely for all liquids. Therefore, we have varied $\tau_{\rm D}G_{\infty}$ around its approximate experimental values (see below) to fit the calculated $c_v$ to the experimental data.

\begin{figure}
\begin{center}
{\scalebox{0.65}{\includegraphics{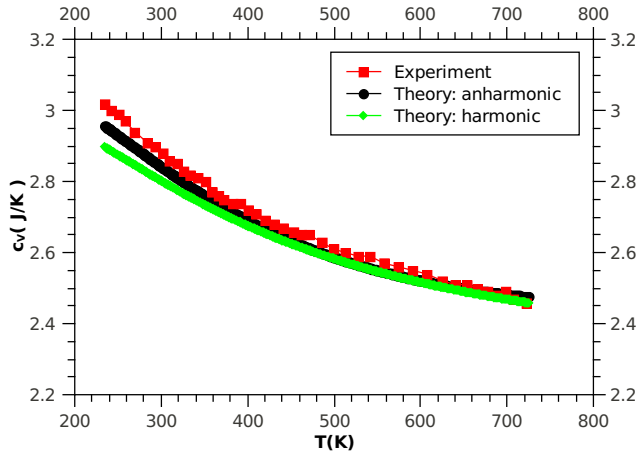}}}
\end{center}
\caption{(Color online). Experimental and theoretical $c_v$ for liquid Hg. Experimental $c_v$ is shown in square symbols. Calculated harmonic and anharmonic $c_v$ are shown in the bottom and thick black curve, respectively. $\tau_{\rm D}G_{\infty}=5.5\cdot 10^{-4}$ Pa$\cdot$s and $6.272\cdot 10^{-4}$ Pa$\cdot$s for theoretical harmonic and anharmonic $c_v$, respectively. $\alpha=1.60\cdot10^{-4}$ K$^{-1}$.}
\label{hg}
\end{figure}

\begin{figure}
\begin{center}
{\scalebox{0.65}{\includegraphics{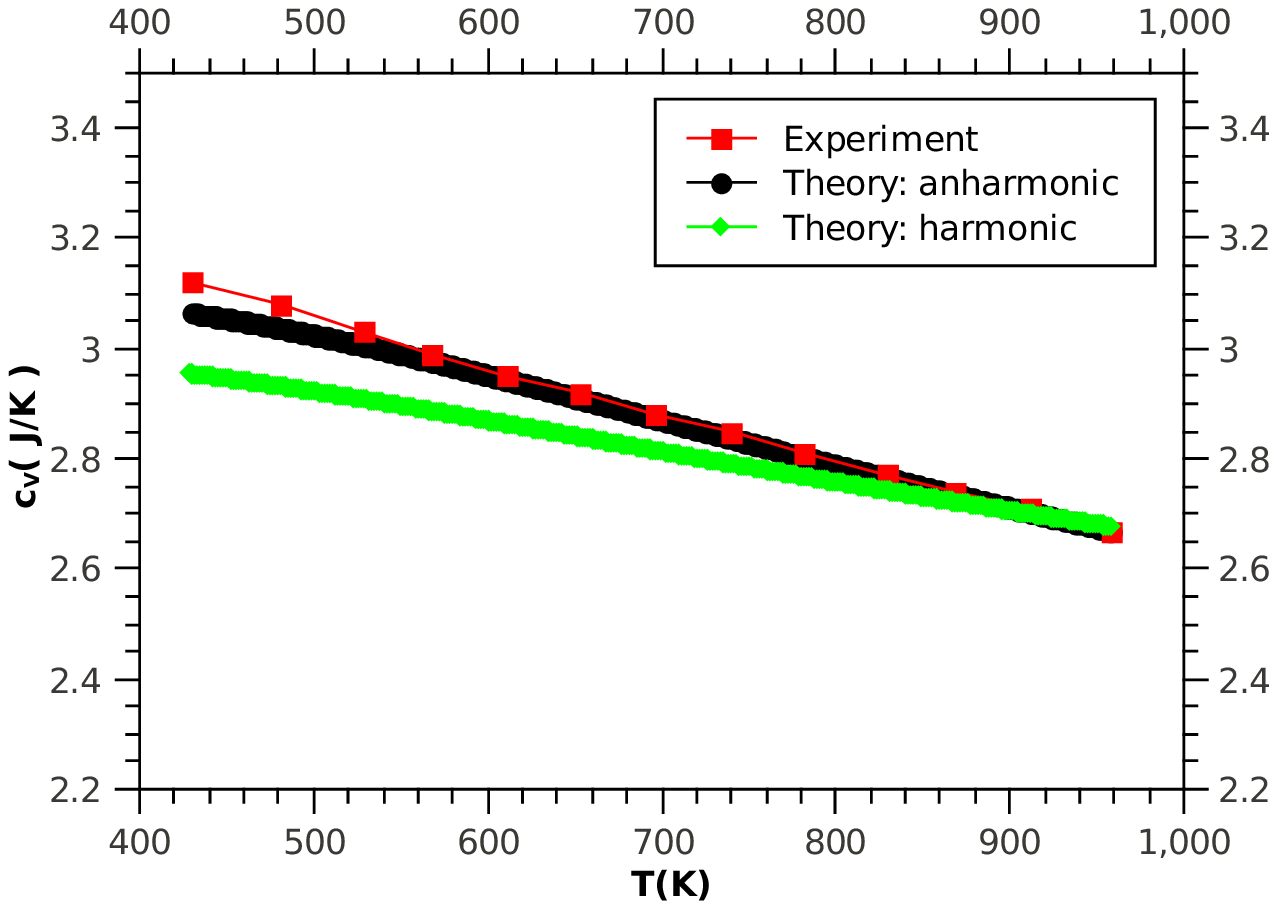}}}
\end{center}
\caption{(Color online). Experimental and theoretical $c_v$ for liquid In. Experimental $c_v$ is shown in square symbols. Calculated harmonic and anharmonic $c_v$ are shown in the bottom and thick black curve, respectively. $\tau_{\rm D}G_{\infty}=4.04\cdot 10^{-4}$ Pa$\cdot$s and $4.88\cdot 10^{-4}$ Pa$\cdot$s for theoretical harmonic and anharmonic $c_v$, respectively. $\alpha=1.22\cdot10^{-4}$ K$^{-1}$.}
\label{in}
\end{figure}

\begin{figure}
\begin{center}
{\scalebox{0.65}{\includegraphics{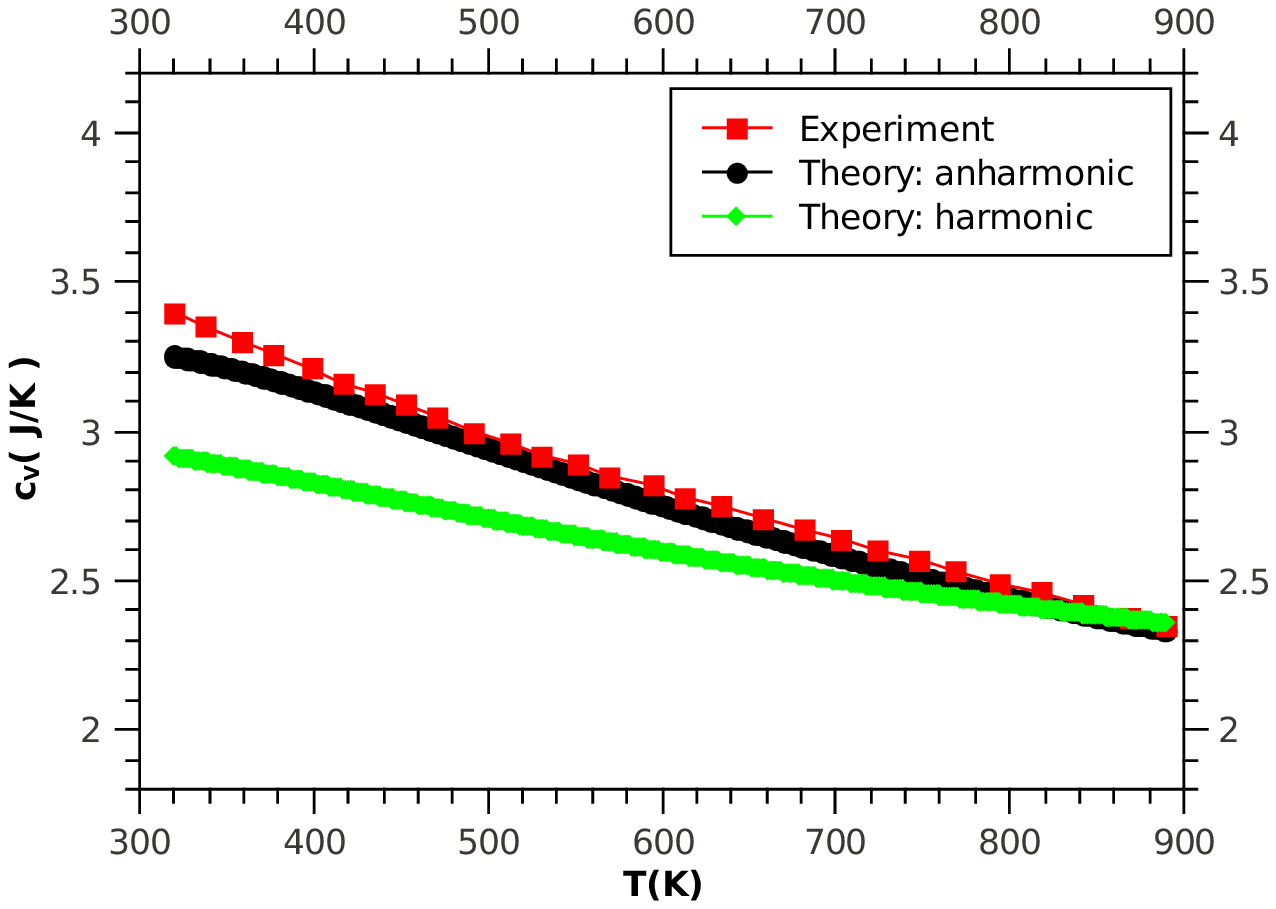}}}
\end{center}
\caption{(Color online). Experimental and theoretical $c_v$ for liquid Cs. Experimental $c_v$ is shown in square symbols. Calculated harmonic and anharmonic $c_v$ are shown in the bottom and thick black curve, respectively. $\tau_{\rm D}G_{\infty}=1.19\cdot 10^{-4}$ Pa$\cdot$s and $2.07\cdot 10^{-4}$ Pa$\cdot$s for theoretical harmonic and anharmonic $c_v$, respectively. $\alpha=4.92\cdot10^{-4}$ K$^{-1}$.}
\label{cs}
\end{figure}

\begin{figure}
\begin{center}
{\scalebox{0.65}{\includegraphics{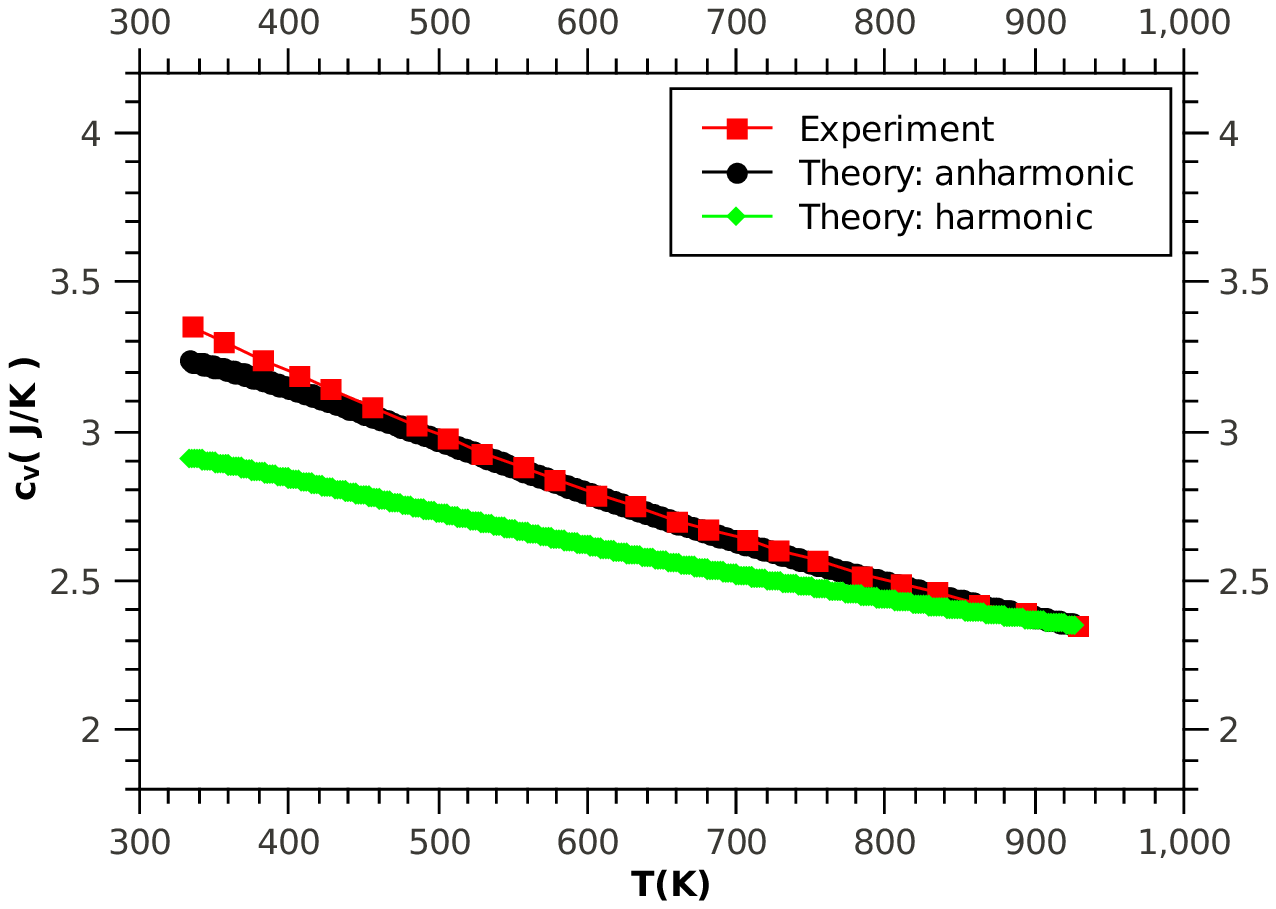}}}
\end{center}
\caption{(Color online). Experimental and theoretical $c_v$ for liquid Rb. Experimental $c_v$ is shown in square symbols. Calculated harmonic and anharmonic $c_v$ are shown in the bottom and thick black curve, respectively. $\tau_{\rm D}G_{\infty}=1.46\cdot 10^{-4}$ Pa$\cdot$s and $1.914\cdot 10^{-4}$ Pa$\cdot$s for theoretical harmonic and anharmonic $c_v$, respectively. $\alpha=4.52\cdot10^{-4}$ K$^{-1}$.
}
\label{rb}
\end{figure}

\begin{figure}
\begin{center}
{\scalebox{0.65}{\includegraphics{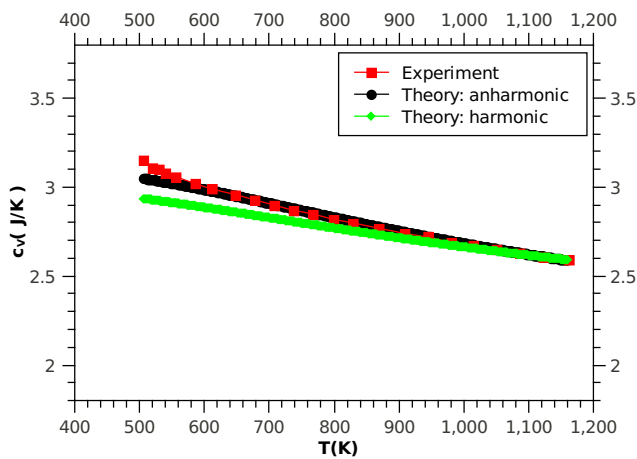}}}
\end{center}
\caption{(Color online). Experimental and theoretical $c_v$ for liquid Sn. Experimental $c_v$ is shown in square symbols. Calculated harmonic and anharmonic $c_v$ are shown in the bottom and thick black curve, respectively. $\tau_{\rm D}G_{\infty}=4.67\cdot 10^{-4}$ Pa$\cdot$s and $5.67\cdot 10^{-4}$ Pa$\cdot$s for theoretical harmonic and anharmonic $c_v$, respectively. $\alpha=1.11\cdot10^{-4}$ K$^{-1}$.}
\label{sn}
\end{figure}

In Figures (\ref{hg}-\ref{sn}), we show the experimental $c_v$ where the electronic contribution was subtracted \cite{grimvall,wallace}. We also show $c_v$, calculated from both harmonic and anharmonic theory using Eqs. (\ref{10},\ref{11}). Overall, Figures (\ref{hg}-\ref{sn}) show reasonably good agreement between the theoretical and experimental $c_v$.

The agreement is somewhat worse at low temperature, however the maximal difference between the predicted and experimental values is comparable with the experimental uncertainty of $c_v$ of 0.1--0.2 J/K \cite{grimvall}. We note that $C_v=3-\left(\frac{\tau_{\rm D}G_\infty}{\eta}\right)^3+\frac{3T}{\eta}\left(\frac{\tau_{\rm D}G_\infty}{\eta}\right)^3\frac{{\rm d}\eta}{{\rm d} T}$ from Eq. (\ref{10}), and similar $\propto\frac{{\rm d}\eta}{{\rm d} T}$ terms appear if Eq. (\ref{11}) is used. Here, the second term is small at low temperature where $\tau\gg\tau_{\rm D}$ and therefore $\eta\gg\tau_{\rm D}G_\infty$. The last term depends on both $\eta$ and $\frac{{\rm d}\eta}{{\rm d} T}$. $\frac{{\rm d}\eta}{{\rm d} T}$ is largest at low temperature, therefore the last term is most sensitive to the fitted slope of $\eta$ at low temperature and its extrapolation, and is expected to contribute most to the difference between the calculated and experimental $c_v$.

Importantly, the theoretical curves in Figures (\ref{hg}-\ref{sn}) are calculated using physically sensible values of parameter $\tau_{\rm D}G_{\infty}$ in the (1-6)$\cdot 10^{-4}$ Pa$\cdot$s range (see captions in Figures (\ref{hg}-\ref{sn})). For example, if $G_{\infty}$ of Hg is about 5 GPa \cite{wallace1}, the used values of $\tau_{\rm D}G_{\infty}$ in Figure (\ref{hg}) imply $\tau_{\rm D}$ of about 0.1 ps, a typical value of Debye vibrational period which is, furthermore, consistent with the experimental value \cite{hgtau0}. Similarly, if $G_{\infty}$ of Cs is 0.77 GPa \cite{cstau0}, the used values of $\tau_{\rm D}G_{\infty}$ in Figure (\ref{cs}) imply $\tau_{\rm D}$ of about 0.1 ps, equally consistent with experiments \cite{cstau0}. We also note that the used value of $\tau_{mt}G_{\infty}$ is close to that in other liquids \cite{pilgrim}.

Similarly to $\tau_{\rm D}G_{\infty}$, the values of $\alpha$ used to calculate the anharmonic $c_v$ (see captions to Figures (\ref{hg}-\ref{sn})) are close to experimental ones. The experimental $\alpha$ for Hg, In, Rb, Cs and Sn are $1.8\cdot10^{-4}$ K$^{-1}$, 1.11$\cdot10^{-4}$ K$^{-1}$, 3$\cdot10^{-4}$ K$^{-1}$, 3$\cdot10^{-4}$ K$^{-1}$ and 0.87$\cdot10^{-4}$ K$^{-1}$, respectively \cite{wallace}, in reasonably good agreement with the values used in Figures (\ref{hg}-\ref{sn}).

We observe that for all five liquids, harmonic $c_v$ is systematically below the experimental value at low temperature. This is not surprising because the maximal low-temperature value of harmonic $c_v$ is 3, according to Eq. (\ref{10}). As Figures (\ref{hg}-\ref{sn}) show, the anharmonic contribution, calculated using Eq. (\ref{11}), is important in order to reproduce the experimental behavior.

In summary, we find that overall, theoretical and experimental values of $c_v$ agree reasonably well, using physically sensible values of parameters $\tau_{\rm D}G_{\infty}$ and $\alpha$.

We note that above we have discussed the behavior of $c_v$. The quantity that is frequently measured in the experiment is the constant-pressure heat capacity, $C_p$. For several low-viscous liquids, $C_p$ weakly depends on temperature, in contrast to the constant-volume heat capacity, $C_v$ \cite{grimvall,md1,pott}. On the other hand, for highly viscous liquids, $C_p$ increases, approximately linearly, with temperature \cite{mckenna}. This behavior can be rationalized in the proposed picture as follows. We re-write the known relationship, $C_p=C_v+VT\alpha^2B$ \cite{landau}, as $c_p=c_v(1+\gamma\alpha T)$, where $c_p=\frac{C_p}{N}$ and $\gamma=\frac{V\alpha B}{C_v}$ is the Gr\"{u}neisen parameter, which is on the order of unity in various systems \cite{anderson}. For highly viscous liquids where $\tau\gg\tau_{\rm D}$, the last term in Eq. (\ref{anharm2}) representing the contribution from the decreasing number of transverse waves, $\left(3-\left(\frac{\omega_0}{\omega_{\rm D}}\right)^3\right)$, can be neglected. Then, $c_v=3(1+\alpha T)$, giving $c_p=3(1+(\gamma+1)\alpha T)$, where we have neglected the square of the small parameter $\alpha T$. This is consistent with the linear increase of experimental $c_p$ \cite{mckenna}. On the other hand, in low-viscous liquids where $\tau$ approaches $\tau_{\rm D}$ at high temperature, the contribution of the decreasing number of transverse modes in Eq. (\ref{anharm2}) can not be neglected. In this case, $c_p=c_v(1+\gamma\alpha T)$ is the product of temperature-decreasing $c_v$ due to the decreasing contribution of transverse waves (see Figures (\ref{hg}-\ref{sn})) and temperature-increasing term, $\propto\alpha T$. As a result, $c_p$ is less sensitive to temperature, as is seen experimentally.

\section{Two approaches to liquid energy: discussion}

It is interesting to compare the proposed approach of calculating liquid energy to previous theories. A traditional way to calculate liquid energy was based on the calculation of liquid potential energy in addition to gas kinetic energy \cite{landau,frenkel,ziman,yip,march,hansen}. In this sense, the approach is from the gas state, in that a liquid is considered to be a gas with interactions switched on. In the approach from the gas state, liquid energy can be written in a generalized form as \cite{landau,frenkel,ziman,yip,march,hansen}:

\begin{equation}
E=\frac{3}{2}NT+\int UF{\rm d}V
\label{gas}
\end{equation}

\noindent where $U$ is the interaction energy between atoms and $F$ is the appropriately normalized correlation function.

In the proposed approach from the solid state, liquid energy is calculated on the basis of solid-like phonons, which therefore accounts for strong interactions from the outset. For convenience and reference to the foregoing discussion, we re-write the harmonic energy, Eq. (\ref{harmo}), as a function of $\tau$:

\begin{equation}
E=NT\left(3-\frac{1}{(\tau\omega_{\rm D})^3}\right)
\label{solid}
\end{equation}

Early approaches based on Eq. (\ref{gas}) were most successfully used for dilute or weakly-interacting systems \cite{landau,frenkel,ziman}. For example, Van Der Waals equation serves to illustrate the effects of interactions and excluded volume. At the same time, $c_v$ of the Van Der Waals liquid is $\frac{3}{2}$, the same as that of the ideal gas \cite{landau}, stressing that this system is far from a real dense liquid.

Historically, the same general approach from the gas state was also used to discuss real dense and strongly-interacting liquids \cite{landau,frenkel,ziman,yip,march,hansen,born,henders,bolm}. We now propose that the approach to liquids from the solid state offers a number of important advantages.

First, experimental liquid $c_v$ changes from 3 right after the melting point to about 2 at high temperature (see Figures (\ref{hg}-\ref{sn})). In the approach from the gas state, the second term in Eq. (\ref{gas}) should be as large as the first one ($\frac{3}{2}NT$) in order to reproduce the low-temperature experimental behavior. On the other hand, the starting value of $c_v=3$ in Eq. (\ref{solid}) is already equal to the experimental one, suggesting that this equation offers a better starting point to discuss liquid $c_v$.

Second, the approach from the gas state does not readily explain the experimental decrease of $c_v$ from 3 to 2. The increase of temperature usually results in broadening of correlation functions, but without the specific form of $F$ at each temperature as well as $U$ and explicit integration, it is not easy to see how Eq. (\ref{gas}) explains the experimental behavior. On the other hand, Eq. (\ref{solid}) explains the experimental data of $c_v$ in a simple and straightforward way.

Third, both harmonic and anharmonic effects in the approach from the gas state are contained in $U$ and $F$, and their separation is not straightforward in general. On the other hand, the anharmonic effects are readily calculated in the approach from the solid state in addition to harmonic terms, from the change of phonon frequencies and thermal expansion (see Eq. (\ref{anharm2})). Consequently, the experimental increase of $c_v$ above the Dulong-Petit value at low temperature is attributed to anharmonicity.

Fourth, it is important to stress that $F$ and $U$ are not generally available for liquids, apart from simple systems such as hard spheres, Lennard-Jones liquids that model Ar, Kr and so on. For these systems, $F$ and $U$ can be determined from experiments or simulations and subsequently used in Eq. (\ref{gas}) to calculate the liquid energy. Unfortunately, neither $F$ nor $U$ are available for liquids with any larger degree of complexity of structure or interactions. For example, many-body correlations \cite{born,henders} and network effects can be strong in familiar liquid systems such as olive oil, SiO$_2$, Se, glycerol or even water \cite{emilio}, resulting in complicated structural correlation functions that can not be reduced to simple two- or even three-body correlations that are often used in Eq. (\ref{gas}). As discussed in Ref. \cite{march}, approximations become difficult to control when the order of correlation functions already exceeds three-body correlations. Similarly, it is challenging to extract multiple correlation functions from the experiment. The same problems exist for interatomic interactions which can be equally multi-body and complex, and consequently not amenable to determination in experiments or simulations.

On the other hand, $\tau$ is available much more widely, and is readily measured in experiments of various types, including dielectric relaxation, NMR and so on. Equally, viscosity $\eta=G_{\infty}\tau$ is routinely measured in liquids using well-known methods, no matter how complicated structural correlations functions or interatomic interactions in a liquid are. Therefore, Eq. (\ref{solid}) can be more readily used to calculate liquid energy from the practical perspective. In future work, we plan to calculate heat capacities of complex liquids in the proposed approach.

Importantly, Eqs. (\ref{gas}) and (\ref{solid}) are related, in that $\tau$ that enters Eq. (\ref{solid}) is itself governed by $F$ and $U$ in Eq. (\ref{gas}): $\tau=\tau(F,U)$. However, this function is generally very complex because it is defined by the activation barriers set by $F$ and $U$ and not by their equilibrium properties. Calculating $\tau$ for a given set of $F$ and $U$ can not be done analytically using a general procedure because the potential energy landscape gets complex even for a small number of particles and possible configurations. This task can be done in a simulation but is equally challenging for realistic systems because $F$ and $U$ are not known except for a small set of simple systems. At the same time, going to the deeper level and calculating the energy from first-principles using $F$ and $U$ is not required in our approach because it is $\tau$ that defines the liquid phonon states from the physical point of view, as proposed by Frenkel originally \cite{frenkel}. Consequently, $F$ and $U$ do not feature in the approach from the solid state (see Eq. (\ref{solid})), just as they do not in the energy of a solid.

Fifth and finally, the proposed approach from the solid state can be said to be more general, or universal, as compared to the approach from the gas state, in the following sense. In the approach from the gas state, the energy strongly depends on $U$ as well as on $F$ (see Eq. (\ref{gas}). It is for this reason that Landau and Lifshitz state the liquid energy strongly depends on the specific form of $U$ and therefore can not be calculated in general form \cite{landau}. Lets now consider liquids with very different structural correlations and interatomic interactions such as, for example, H$_2$O, Hg, AsS, olive oil and glycerol. Even if their energy could be calculated on the basis of Eq. (\ref{gas}), large differences in $U$ and $F$ would imply different values of energy, in agreement with the argument of Landau and Lifshitz \cite{landau}. On the other hand, as long as $\tau$ of the above liquids is the same at certain temperature (different for each liquid), Eq. (\ref{solid}) predicts that their energy is the same. In this sense, expressing the liquid energy as a function of $\tau$ only is a more general description because $\tau$ is common to all liquids whereas $F$ and $U$ are system-specific as in the above examples.

\section{Conclusions}

In summary, we have developed the approach to liquids from the solid state by incorporating the effects of anharmonicity and thermal expansion. In this approach, we have found a good agreement between theoretical predictions and experimental data for five commonly discussed liquids. We have proposed that the approach to liquids from the solid state offers a number of advantages as compared to the previous approach from the gas state.

We are grateful to V. V. Brazhkin for discussions and to SEPnet, EPSRC, Myerscough Bequest and the School of Physics in Queen Mary University for support.

\end{document}